\documentclass[aps,prd,twocolumn,superscriptaddress,amsmath,amssymb,floatfix,longbibliography,10pt]{revtex4-1}

\usepackage{graphicx}
\usepackage{amssymb,amsmath,mathtools}
\usepackage{bbm}
\usepackage{color}
\usepackage[colorlinks,urlcolor=blue,linkcolor=blue,anchorcolor=blue,citecolor=blue]{hyperref}

\graphicspath{{./}{./plots/}}

\newcommand{\nccp}{NCCP$^1$}

\newcommand{\qedgn}{QED$_3$-GN}

\begin{document}

\title{Critical behavior of the \texorpdfstring{QED$_3$}{QED3}--Gross-Neveu model: Duality and deconfined criticality}

\author{Lukas Janssen}
\affiliation{Institut f\"ur Theoretische Physik, Technische Universit\"at Dresden, 01062 Dresden, Germany}
\author{Yin-Chen He}
\affiliation{Department of Physics, Harvard University, Cambridge, Massachusetts 02138, USA}


\begin{abstract}
We study the critical properties of the QED$_3$--Gross-Neveu model with $2N$ flavors of two-component Dirac fermions coupled to a massless scalar field and a U(1) gauge field. For $N=1$, this theory has recently been suggested to be dual to the SU(2) noncompact CP$^1$  model that describes the deconfined phase transition between the  N\'eel antiferromagnet and the valence bond solid on the square lattice. For $N=2$, the theory has been proposed as an effective description of a deconfined critical point between chiral and Dirac spin liquid phases, and may potentially be realizable in spin-$1/2$ systems on the kagome lattice. We demonstrate the existence of a stable quantum critical point in the QED$_3$--Gross-Neveu model for all values of $N$. This quantum critical point is shown to escape the notorious fixed-point annihilation mechanism that renders plain QED$_3$ (without scalar-field coupling) unstable at low values of $N$. The theory exhibits an upper critical space-time dimension of four, enabling us to access the critical behavior in a controlled expansion in the small parameter $\epsilon = 4-D$. We compute the scalar-field anomalous dimension~$\eta_\phi$, the correlation-length exponent~$\nu$, as well as the scaling dimension of the flavor-symmetry-breaking bilinear $\bar\psi\sigma^z\psi$ at the critical point, and compare our leading-order estimates with predictions of the conjectured duality.
\end{abstract}

\maketitle


\section{Introduction}

At zero temperature, strongly-correlated systems exhibit transitions between different phases of matter upon tuning non-temperature parameters, such as external pressure or chemical doping.
Just as their classical counterparts, these quantum phase transitions are characterized by only a few universal properties that are governed by an associated continuum quantum field theory~\cite{sachdevbook}.
Most quantum phase transitions have a classical analog and can be characterized in terms of a local order parameter that allows to classify and distinguish different phases of matter---a property that is commonly referred to as Landau's symmetry breaking paradigm.

There exist, however, exotic phase transitions which are inherently quantum mechanical and for which the Landau theory is inapplicable. 
The most familiar example is the putative deconfined quantum critical point between two different symmetry-breaking phases of a spin-$1/2$ system on the square lattice---the N\'eel and valence bond solid (VBS) states~\cite{senthil2004a, senthil2004b}. 
The deconfined critical point is characterized by fractionalized bosonic spinons on the complex projective space CP$^1$ coupled to an emergent noncompact $\mathrm{U}(1)$ gauge field. These degrees of freedom emerge only directly at the critical point, but are ``confined'' in either phase. The appropriate theoretical description of the criticality is given by a strongly interacting gauge field theory---the noncompact CP$^1$ (\nccp)  model.
More recently, new types of such non-Landau transitions have been suggested, which are similarly governed by strongly interacting gauge theories~\cite{Grover2013,Lu2014,he2015b}. This includes transitions between different long-range entangled phases, such as the Dirac and chiral spin liquid phases~\cite{Kalmeyer1987,Wen1989, Hastings2000}, between short-range entangled phases, e.g., symmetry-protected topological phases~\cite{Chen2013}, and between phases with anticommuting fermion mass terms~\cite{sato2017}.

The strongly interacting gauge theories that describe the above deconfined critical points are also of wide fundamental interest with respect to various duality webs that were proposed recently. Via these dualities, several seemingly different theories can be mapped onto each other and themselves. The easy-plane version of the \nccp\ model, for instance, has been argued to be self-dual \cite{motrunich2004, senthil2004b}, which can be understood as a consequence of the well-known bosonic particle-vortex duality \cite{peskin1978, dasgupta1981, fisher1989}. 
Specifically, the spinon field content of the \nccp\ model can be viewed as either two flavors of bosonic spinons or two flavors of bosonic vortices. 
Building on the Dirac theory of the half-filled Landau level~\cite{son2015}, several works suggest a fermionic counterpart of the particle-vortex duality~\cite{wang2015,mross2016,metlitski2016}. This has lead to a number of fascinating novel duality conjectures, including ones that relate purely bosonic systems to fermionic theories~\cite{xu2015,seiberg2016,karch2016,murugan2017,hsien2016,Mross2017,Chen2017}.
Early proposals of a duality between the easy-plane \nccp\ model and quantum electrodynamics in 2+1 dimensions (QED$_3$)~\cite{alicea2005,senthil2006} have recently undergone various consistency checks, corroborating the intimate relationship between these seemingly different theories~\cite{karch2016,wang2017,Qin2017}. In a similar way, the SU(2) invariant \nccp\ model has been argued to be self-dual as well as to be dual to QED$_3$ coupled to a critical real scalar field---a theory that was coined ``QED$_3$--Gross-Neveu'' (\qedgn) model \cite{wang2017}. 
An immediate consequence of this conjectured duality is the emergence of an enlarged SO(5) symmetry, which was numerically observed earlier~\cite{nahum2015a, nahum2015b}. 

While the infrared fate of QED$_3$ has extensively been discussed in the last three decades~\cite{appelquist1988, kubota2001, hands2004, braun2014, raviv2014, janssen2016, herbut2016, dipietro2016, gusynin2016, kotikov2016b, chester2016, karthik2016a, karthik2016b}, the infrared structure of the \qedgn\ model has, to the best of our knowledge, not been studied before. In this work, we demonstrate that the \qedgn\ model exhibits a stable fixed point of the renormalization group (RG) for all fermion flavor numbers $N$. In particular, we demonstrate that the coupling to the critical scalar field prevents the mechanism of fixed-point annihilation that is responsible for the instability of plain QED$_3$ at low values of $N$ \cite{kubota2001, braun2014, janssen2016, herbut2016}. The stable fixed point can be approached by tuning a single parameter, such as the scalar-field mass, and thus can be associated with a continuous quantum phase transition. The existence of this quantum critical point for two flavors of two-component fermions is a necessary condition for the \nccp--\qedgn\ duality to hold. We compute the critical exponents $\eta_\phi$ (order-parameter anomalous dimension) and $\nu$ (correlation-length exponent) as well as the scaling dimension of the flavor-symmetry-breaking bilinear $\bar\psi\sigma^z\psi$, within an $\epsilon$ expansion around the upper critical space-time dimension of four.
If the duality holds, the universal exponents at this quantum critical point in the physical space-time dimension of $D=2+1$ can be uniquely mapped onto those of the SU(2) invariant \nccp\ model, and we compare our leading-order estimates with numerical results for the bosonic systems \cite{sandvik2007,melko2008,nahum2015a}.
Our work represents the first step towards a proper quantification of the critical behavior of the \qedgn\ model. In the plain Gross-Neveu system (without the coupling to the gauge field), significant progress was made previously by employing high-order $\epsilon$ expansion \cite{gracey2016, mihaila2017}, the functional renormalization group \cite{janssen2014, heilmann2015, knorr2016}, the conformal bootstrap approach \cite{iliesu2016, iliesu2017}, and sign-free quantum Monte Carlo simulations \cite{assaad2013,chandrasekharan2013,toldin2015,li2015,otsuka2016}. Extending these advances to the \qedgn\ case, and comparing with results for the \nccp\ model, should allow to prove or disprove the duality conjecture in future studies. 

The critical behavior of the \qedgn\ model is of interest for yet another reason: This model for the case with four two-component fermion flavors has recently been suggested to describe the deconfined critical point between the chiral spin liquid and the U(1) Dirac spin liquid phases~\cite{he2015b}. Both phases, and their transition, are potentially realizable in spin-$1/2$ systems on the kagome lattice~\cite{he2014, Gong2014, he2015a, he2016}. Our finding of a stable fixed point corroborates this proposal, and the predictions for the critical behavior may facilitate a numerical test of it in the future.

The paper is organized as follows: In the following section, we define the \qedgn\ theory and review the proposed dualities and the potential applicability to deconfined criticality. In Sec.~\ref{sec:fermionic}, we compute the RG flow in a fermionic language that allows to make contact with previous works on the plain QED$_3$ theory. The $4-\epsilon$ expansion of the \qedgn\ theory is performed in Sec.~\ref{sec:eps-expansion}. In Sec.~\ref{sec:conclusions}, we summarize our results and attempt some conclusions in light of the conjectured \nccp--\qedgn\ duality.


\section{Model} \label{sec:model}

We are interested in the \qedgn\ theory, defined by the Lagrangian
\begin{align} \label{eq:lagrangian}
\mathcal L_{\psi\phi} & =  \bar \psi_i \left[\gamma_\mu (\partial_\mu - i  a_\mu)\right] \psi_i
+ \frac{1}{2e^2}(\epsilon_{\mu\nu\rho}\partial_\nu a_\rho)^2
\nonumber \allowdisplaybreaks[0] \\ & \quad
+ g\phi \bar \psi_i \psi_i 
+ \frac{1}{2}\phi(r - \partial_\mu\partial_\mu) \phi 
+ \lambda \phi^4,
\end{align}
in $D=2+1$ Euclidean space-time dimensions. The summation convention over repeated indices is assumed. We consider an even number $2N$ of two-component Dirac fermion flavors $\psi_i$ and $\bar\psi_i$, $i=1,\dots,2N$. The parity symmetry is therefore explicitly preserved for any integer $N$ and the flavor symmetry is $\mathrm{U}(2N)$. The $2\times 2$ Dirac matrices $\gamma_\mu$ fulfill the Clifford algebra $\{\gamma_\mu, \gamma_\nu\} = 2\delta_{\mu\nu} \mathbbm{1}_2$, with $\mu,\nu = 0,1,2$. The fermions couple to the $\mathrm{U}(1)$ gauge field $a_\mu$ with charge $e^2$. The explicit calculations presented below are performed in a general $R_\xi$ gauge with undetermined gauge-fixing parameter $\xi$, by adding $\mathcal L_\text{gf} = -\frac{1}{2\xi} (\partial_\mu a_\mu)^2$ to the Lagrangian. This enables us to verify the gauge independence of our results. $\phi$ is a real scalar field that is odd under the time-reversal symmetry (TRS). It interacts with the fermions through the Yukawa coupling $g$, and with itself through the $\phi^4$ coupling $\lambda$. $r$ is a tuning parameter for the TRS breaking transition, indicated by the formation of a finite scalar-field expectation value, $\langle \phi \rangle \neq 0$.
As mentioned in the introduction, this \qedgn\ theory has interesting applications:

(1) By applying the boson-fermion duality~\cite{Chen1993,karch2016,seiberg2016}, and building on earlier observations~\cite{senthil2006}, the case $N=1$ has recently been conjectured to be dual to the bosonic \nccp\ theory \cite{wang2017},
\begin{align}\label{eq:NCCP1}
\mathcal L_{z} & = 
\sum_{\alpha=1,2} |(\partial_\mu -i b_\mu) z_\alpha|^2 
+ \kappa(\epsilon_{\mu\nu\rho}\partial_\nu b_\rho)^2 
\nonumber \\ & \quad 
+ \lambda_0 \left(|z_1|^2+|z_2|^2\right)^2+\lambda_1 |z_1|^2 |z_2|^2.
\end{align}
Here, $z = (z_1,z_2)$ are complex bosonic fields and $b_\mu$ is a $\mathrm U(1)$ gauge field.
%
%
When $\lambda_1 = 0$, the theory has an explicit $\mathrm{SU}(2)$ symmetry. We will refer to this case as $\mathrm{SU}(2)$ \nccp\ model. This theory is believed to describe the deconfined critical point between the N\'eel and VBS phases on the square lattice~\cite{senthil2004a,senthil2004b}. For $\lambda_1 \neq 0$, the theory has an easy-plane anisotropy with a residual $\mathrm{O}(2)$ symmetry and is relevant for spin models with an XY symmetry.

The postulated dualities between Eq.~\eqref{eq:lagrangian} and Eq.~\eqref{eq:NCCP1} are as follows: 
(i) The plain QED$_3$ theory with the scalar field $\phi$ decoupled (formally corresponding to the limit of large tuning parameter $r$) is dual to the easy-plane \nccp\ model with $\lambda_1 \neq 0$. 
(ii) The critical \qedgn\ theory with $r$ tuned such that $\phi$ becomes gapless is dual to the $\mathrm{SU}(2)$ \nccp\ model with $\lambda_1 = 0$. 
While these proposed dualities have passed a number of consistency checks~\cite{wang2017}, we should emphasize that, at present, they lack any formal or numerical proof and should be considered as conjectural.
The conjectures, however, do predict a number of nontrivial relations between the universal exponents that describe the critical behaviors of these theories, allowing in principle to verify or falsify the conjectures on a quantitative level. For the case (ii), the scalar field $\phi$ is identified with $z^\dagger \sigma^z z$, which is an element of the N\'eel-VBS SO(5) order parameter. The scalar anomalous dimension $\eta_\phi$ in the \qedgn\ theory should therefore coincide with the anomalous dimensions $\eta_\text{N\'eel}$ and $\eta_\text{VBS}$ in the spin systems. Furthermore, the dual to the $\phi^2$ operator corresponds to a rank-2 tensor representation of the SO(5) critical theory. 
The latter contains the operator $z^\dagger z$ that tunes through the N\'eel-VBS transition, and therefore the correlation-length exponents $\nu_\text{\qedgn}$ and $\nu_\text{N\'eel-VBS}$ in the two systems should also coincide. Another consequence of the proposed duality is that $z^\dagger z$ can also be identified with the flavor-symmetry breaking fermion bilinear $\bar\psi\sigma^z\psi$. Therefore, the scaling dimension of $\bar\psi\sigma^z\psi$ should also coincide with the scaling dimension of $\phi^2$. This statement is particularly interesting, because it allows a nontrivial test of the duality conjecture fully within the \qedgn\ theory---without the need to compare with a different system. A similar relation between $\eta_\phi$ and the scaling dimensions of certain monopole operators also follows~\cite{wang2017}.

Obviously, a necessary condition for such a duality to hold is the existence of a stable interacting RG fixed point.
In fact, for the case of plain QED$_3$, the emergence of conformal invariance at low energy, and therewith the existence of a conformal fixed point, can be established when the number of fermions $N$ is large~\cite{jackiw1981,appelquist1988}. At low values of $N$, however, a generic mechanism that may destabilize the conformal fixed point is the collision and subsequent annihilation with another, quantum critical, fixed point~\cite{kubota2001,braun2014,janssen2016,herbut2016}, very much like in the case of the Abelian Higgs model~\cite{halperin1974, nahum2015b} as well as a number of further examples~\cite{jaeckel2006,kaplan2009,herbut2014,janssen2017}. Such instability is driven by strong gauge fluctuations and is therefore ubiquitous in asymptotically safe gauge theories~\cite{kaplan2009}. It leads to an essential singularity of physical observables at a critical flavor number $N_\mathrm{c}$, below which the conformal state becomes unstable. The actual value of $N_\mathrm{c}$ in QED$_3$, however, and with it the important question whether $N_\mathrm{c}$ is above~\cite{appelquist1988, kubota2001, hands2004, braun2014, raviv2014, herbut2016, dipietro2016, gusynin2016, kotikov2016b} or below~\cite{karthik2016a,karthik2016b,chester2016} the physically relevant values, as well as the nature of the low-$N$ phase~\cite{janssen2016}, has been a matter of intense debate within the last three decades.
Similarly, on the bosonic side of the duality, the question whether the transition in the easy-plane \nccp\ model is intrinsically first order, or if a lattice model that hosts a continuous transition between the XY antiferromagnetic and VBS states can be constructed, has been controversially discussed in the past~\cite{kragset2006,emidio2016,emidio2017}. Some very recent numerical studies suggest a continuous transition~\cite{Qin2017,Zhang2017}, with exponents that are potentially in agreement with the conformal phase of plain QED$_3$~\cite{karthik2016b}.

By contrast, the infrared behavior of the \qedgn\ model has, to the best of our knowledge, not been investigated before~\cite{roy2013}. In the next section, we demonstrate that the coupling to the critical scalar field $\phi$ in fact stabilizes the theory, despite the presence of strong gauge fluctuations, and it leads to a stable fixed point that governs a continuous transition into a state with spontaneously broken TRS.

(2) The case $N=2$ is relevant for the physics of spin liquid states on the kagome lattice. Despite tremendous efforts, the actual nature of the quantum ground state of the Heisenberg antiferromagnet on the kagome lattice to date has not been established beyond doubt. The most promising candidates are either a gapped $\mathbbm{Z}_2$ spin liquid~\cite{yan2011,depenbrock2012,jiang2012} or a gapless $\mathrm{U}(1)$ Dirac spin liquid~\cite{ran2007,iqbal2015,he2016}. 
Longer-range spin interactions appear to stabilize yet another spin liquid phase, which is characterized by spontaneous breaking of time reversal symmetry---a chiral spin liquid with anyonic spinon statistics~\cite{he2014, Gong2014}. The transition into this state appears to be continuous~\cite{he2015a}, and if the ground state in the nearest-neighbor model is a Dirac spin liquid, the effective field theory that describes this transition would be the \qedgn\ model with $2N=4$ flavors of two-component fermion flavors~\cite{he2015b}. Determining the critical behavior of the \qedgn\ model may therefore allow to prove or disprove this scenario if the critical behavior of the spin-liquid transition on the kagome lattice becomes possible to be quantified numerically.
%


\section{\texorpdfstring{\qedgn}{QED3-GN} quantum critical point in fermionic RG} \label{sec:fermionic}

The presumed quantum critical point in the theory defined by Eq.~\eqref{eq:lagrangian} with $r$ tuned to criticality demarcates the ordered phase in which the TRS is spontaneously broken, $\langle \phi \rangle \neq 0$, from the time-reversal-symmetric phase, $\langle \phi \rangle = 0$. The infrared behavior of the latter phase is governed by the conformal fixed point of plain QED$_3$, which albeit in turn may be destabilized for low values of $N$ by a collision with another fixed point~\cite{kubota2001,braun2014,janssen2016,herbut2016}. 
In this section, we demonstrate that a critical point that can be identified with the TRS-breaking transition exists for all $N$. In particular, it survives when the conformal fixed point of plain QED$_3$ collides and annihilates with another fixed point when lowering $N$.

\begin{figure*}[t]
\includegraphics[scale=.57]{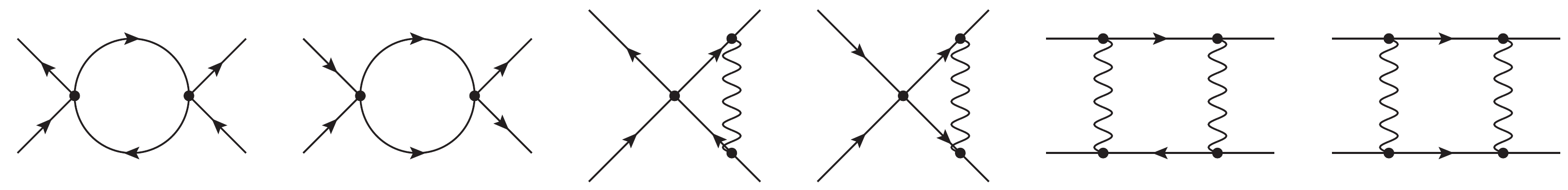}
\caption{One-loop diagrams determining the flows of the four-fermion couplings $u$ and $v$. Solid (wiggly) lines correspond to fermion (gauge) fields.}
\label{fig:fermion}
\end{figure*}

\begin{figure}[b]
\includegraphics[scale=.57]{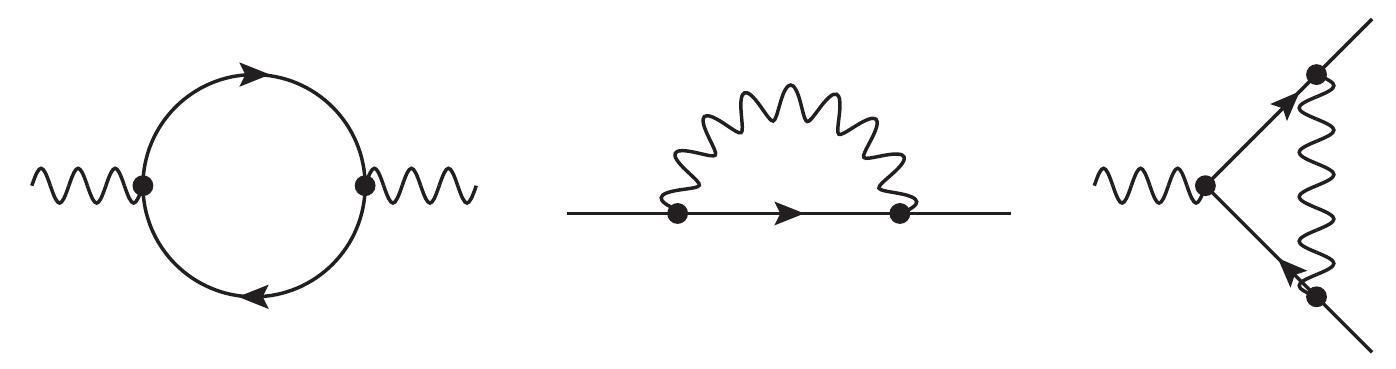}
\caption{Diagrams that determine the gauge-field anomalous dimension $\eta_a$ (left) and the fermion selfenergy (middle). In the flow equation for $e^2$, the contributions from the fermion selfenergy and explicit vertex correction (right) cancel due to the Ward identity associated with the U(1) gauge symmetry.}
\label{fig:gauge}
\end{figure}

In order to make contact with the conformal phase of QED$_3$, we approach the TRS-breaking transition from the symmetric side, $\langle\phi\rangle = 0$. On this side, we may neglect the quartic coupling $\lambda \phi^4$ for simplicity and integrate out $\phi$. This way, we obtain a Gross-Neveu-type four-fermion interaction 
\begin{align}
u \left(\bar \psi_i \psi_i\right)^2, \quad i = 1,\dots, 2N,
\end{align}
with negative four-fermion coupling $u < 0$.
In a RG picture, the transition towards the TRS-breaking state would in this formulation be indicated by an instability of the flow towards divergent $u \to -\infty$ at a finite RG scale.
Once radiative corrections are taken into account, further terms that are not present in the initial action may be generated by the RG. However, symmetry strongly restricts the number of possible terms. On the level of four-fermion interactions, the only term that is compatible with the $\mathrm U(2N)$ flavor symmetry is the Thirring interaction \cite{gies2010},
\begin{align}
v \left( \bar\psi_i \gamma_\mu \psi_i \right)^2.
\end{align}
A minimal low-energy effective theory is therefore given by a $\mathrm U(1)$ gauge theory with $2N$ flavors of two-component Dirac fermions, augmented with Gross-Neveu and Thirring four-fermion interactions:
\begin{align}
 \mathcal L_\psi = \bar \psi_i \gamma_\mu (\partial_\mu - i a_\mu) \psi_i
  + u \left(\bar \psi_i \psi_i\right)^2
  + v \left(\bar \psi_i \gamma_\mu \psi_i \right)^2.
\end{align}
Integrating over the momentum shell from $\Lambda$ to $\Lambda/b$ with $b>1$, the RG flow of this theory reads, to the one-loop order
\begin{align} 
\label{eq:beta-e2}
\frac{d e^2}{d \ln b} & = (1 - \eta_a) e^2, \\
\label{eq:beta-g1}
\frac{d u}{d \ln b} & = - u + \frac{16}{3} e^2 u + 8 e^2 v + 2 e^4 - 4(2N-1) u^2  \nonumber \\
& \quad + 8 v^2 + 12 u v, \\
\label{eq:beta-g2}
\frac{d v}{d \ln b} & = - v + \frac{8}{3} e^2 u + \frac{4}{3}(2N+1) v^2 + 4 u v,
\end{align}
with the gauge-field anomalous dimension $\eta_a = \frac{4}{3}N e^2$. In order to arrive at the above beta functions, we have rescaled the couplings as $e^2/(2\pi^2 \Lambda) \mapsto e^2$, $\Lambda u/(2\pi^2) \mapsto u$, and $\Lambda v/(2\pi^2) \mapsto v$. The corresponding diagrams are depicted in Figs.~\ref{fig:fermion} and \ref{fig:gauge}. The above equations are consistent with previously published ones in the respective limit~\cite{janssen2016}.

\begin{figure*}
\includegraphics[width=0.85\textwidth]{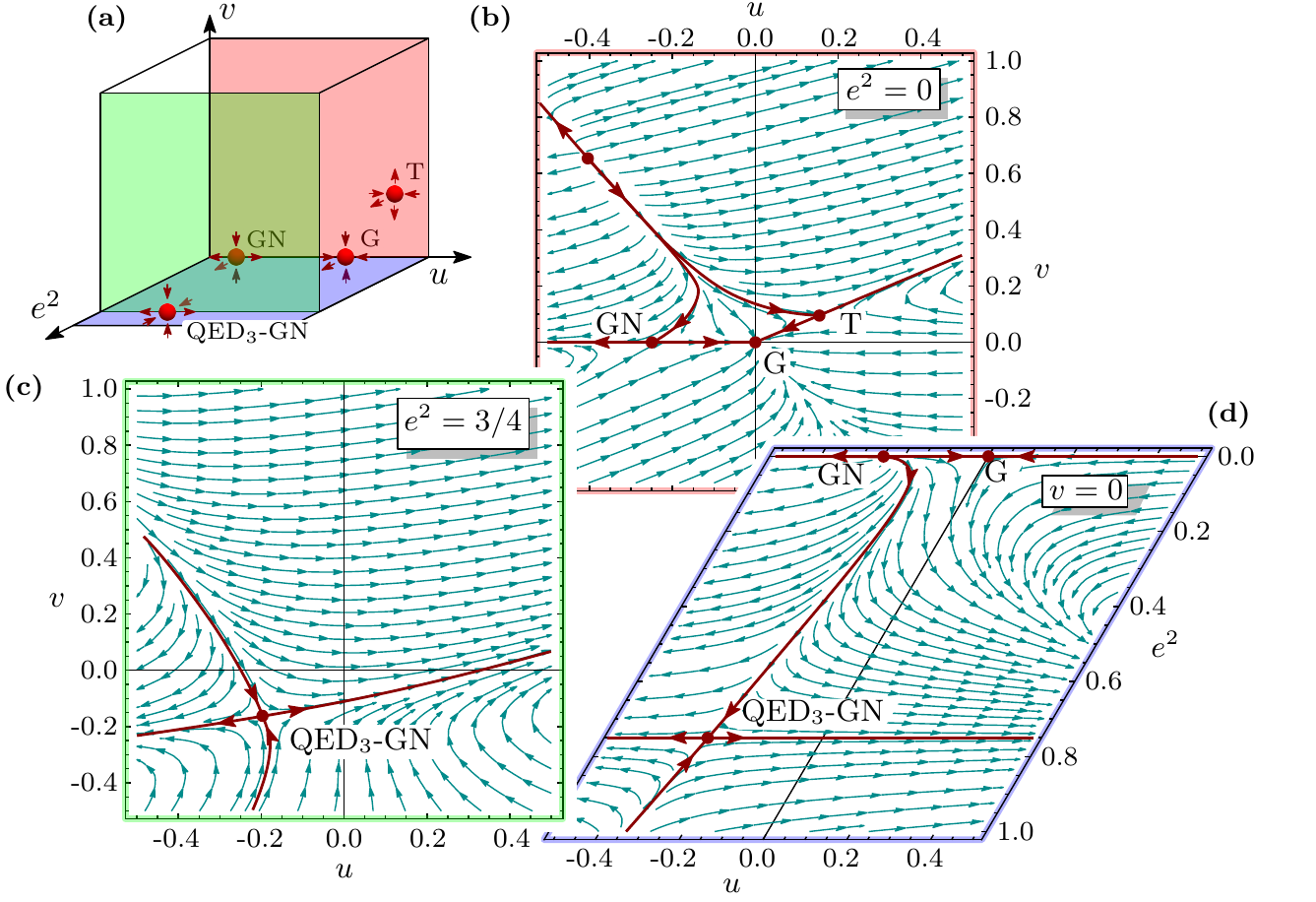}
\caption{RG flow for $N=1$. The panels in (b)--(d) display the RG flow within different subspaces of the full theory space spanned by $u$, $v$, and $e^2$, as schematically depicted in (a). (b) $u$-$v$ plane for $e^2 = 0$. (c) $u$-$v$ plane for $e^2 = e^2_* = 3/4$. (d) $u$-$e^2$ plane for $v = 0$. Besides the Gaussian fixed point (G), the only quantum critical point with just one relevant direction is the \qedgn\ fixed point. It describes the TRS-breaking transition. There are furthermore two critical points in the uncharged sector $e^2 = 0$, which, however, receive a second relevant direction along the $e^2$ axis: the Gross-Neveu fixed point (GN) and the Thirring fixed point (T).} 
 \label{fig:RG-flow}
\end{figure*}

At large $N$, the fixed-point structure can be elucidated analytically. At zero charge $e^2 = 0$, there are two critical fixed points: the Gross-Neveu fixed point (GN), located at
\begin{align}
\text{GN} &: & (e^2_*, u_*, v_*) & = \left(0, -\frac{1}{8N}, 0\right) + \mathcal O(1/N^2),
\end{align}
and the Thrirring fixed point (T) at
\begin{align}
\text{T} &: & (e^2_*, u_*, v_*) & = \left(0, 0, \frac{3}{8N}\right) + \mathcal O(1/N^2).
\end{align}
They are believed to be of relevance in the context of interacting fermions on the honeycomb lattice~\cite{herbut2006, herbut2009, assaad2013, janssen2014}, and have been extensively studied in the past~\cite{braun2011, gracey2016, mihaila2017, gies2010, janssen2012, christofi2007, hands2016,wellegehausen2017}.
The charge $e^2$, however, is RG relevant towards the infrared and flows to a finite fixed-point value $e^2_* = \frac{3}{4N} + \mathcal O(1/N^2)$. In this ``charged'' plane, there are two quantum critical points when $N$ is large. We find
\begin{align} \label{eq:qedgn-fermionic}
 \text{\qedgn}&: & (e^2_*, u_*, v_*) & = \left(\frac{3}{4N}, -\frac{1}{8N}, 0\right) + \mathcal O(1/N^2),
\end{align}
and
\begin{align} \label{eq:g-T}
 \text{g-T}&: & (e^2_*, u_*, v_*) & = \left(\frac{3}{4N}, 0, \frac{3}{8N} \right) + \mathcal O(1/N^2),
\end{align}
and both have precisely one RG relevant direction in the $(e^2, u, v)$ space of couplings. The relevant direction of the former fixed point (``\qedgn'') is aligned along
\begin{align} \label{eq:eigenvector-qedgn}
(e^2, u, v) & = (0, -1, 0) + \mathcal O(1/N),
\end{align}
and therefore describes a transition into a state with $u \to -\infty$. This corresponds to the spontaneous breaking of TRS, and the fixed point should therefore be understood as the projection of the critical point in the full \qedgn\ theory onto the four-fermion coupling space. 
The fixed point in Eq.~\eqref{eq:g-T} (``g-T'') represents a gauged version of the Thirring fixed point. Moreover, we also rediscover~\cite{kubota2001,braun2014,dipietro2016,janssen2016,herbut2016} the fully infrared attractive fixed point that describes the conformal phase of QED$_3$, which in the limit of large $N$ is located at 
\begin{align}
 \text{c-QED$_3$}&: & (e^2_*, u_*, v_*) & = \left(\frac{3}{4N}, 0, 0 \right) + \mathcal O(1/N^2).
\end{align}
By evaluating the fixed-point equations at finite $N$ numerically, we find that it is the g-T fixed point (and not the \qedgn\ fixed point) that approaches c-QED$_3$ and eventually collides and annihilates with the latter at a critical flavor number $N_{\mathrm c}$.
This is in agreement with the previous RG studies~\cite{kubota2001,braun2014,janssen2016,herbut2016}. In our simple approximation, this fixed-point annihilation happens at $N_\mathrm{c} \approx 6$, a number which should be expected to receive corrections when going beyond the present one-loop order. In any case, the point we would like to emphasize here is that the \qedgn\ fixed point, in contrast to the c-QED$_3$ and g-T fixed points, \emph{survives} across the transition at $N_\mathrm{c}$ and continuous to exist for all values of $N$.
For the case of $N=1$, relevant to the duality conjecture, the RG flow in the coupling space spanned by $e^2$, $u$, and $v$ is illustrated in Fig.~\ref{fig:RG-flow}, showing the fixed points GN and T in the uncharged sector $e^2 = 0$ and the quantum critical \qedgn\ fixed point in the RG attractive plane $e^2 = e_*^2$.

\begin{figure*}[t]
\includegraphics[scale=.57]{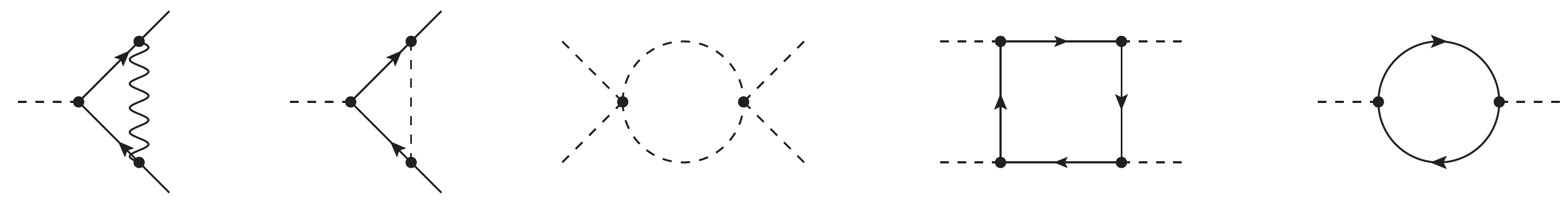}
\caption{One-loop diagrams determining the flows of the Yukawa coupling $g^2$ and the $\phi^4$ coupling $\lambda$, as well as the scalar-field anomalous dimension $\eta_\phi$. Dashed lines correspond to the scalar field.}
\label{fig:scalar}
\end{figure*}

From the flow of the relevant direction [Eq.~\eqref{eq:eigenvector-qedgn}], we obtain the correlation-length exponent at the \qedgn\ fixed point as
\begin{align} \label{eq:nu-fermionic}
1/\nu = 1 + \mathcal O(1/N).
\end{align}
In the above equation, we have displayed only the leading-order value within the $1/N$ expansion, for which our one-loop flow equations are sufficient~\cite{kaveh2005}. The computation of the $1/N$ correction requires the knowledge of the two-loop flow. This is left for future work.
The scaling dimension of the TRS-breaking fermion bilinear $\bar\psi\psi$ can be determined by computing the flow of a small symmetry-breaking perturbation of the form $\Delta \bar\psi\psi$. At the one-loop order~\cite{janssen2016},
\begin{align} \label{eq:flow-Delta}
\frac{d \Delta}{d \ln b} = \left(1 - 2(4N-1)u + 6 v + \frac{8}{3}e^2\right) \Delta + \mathcal O(\Delta^2).
\end{align}
Therewith, we find
\begin{align}
[\bar\psi\psi]_\text{\qedgn} = 3 - [\Delta] = 1 + \mathcal O(1/N)
\end{align}
at the \qedgn\ fixed point, corresponding to an anomalous dimension 
\begin{align} \label{eq:eta-fermionic}
\eta_\phi = 1 + \mathcal O(1/N)
\end{align}
of the TRS order parameter $\langle\phi\rangle \propto \langle\bar\psi\psi\rangle$. Note that the scaling dimensions at the other critical fixed points, such as g-T and T, would be $[\bar\psi\psi]_\text{g-T} = [\bar\psi\psi]_\text{T} = 2 + \mathcal O(1/N)$, and thus these fixed points are, pictorially speaking, ``less unstable'' towards the TRS-breaking perturbation.
Along the same line, we can obtain the scaling dimension of the flavor-symmetry breaking bilinear $\bar\psi\sigma^z\psi \equiv \bar\psi_i(\sigma^z \otimes \mathbbm 1_{N})_{ij}\psi_j$. At the \qedgn\ fixed point, it becomes
\begin{align}
[\bar\psi \sigma^z \psi]_\text{\qedgn} = 2 + \mathcal O(1/N).
\end{align}
This corroborates our conclusion that the fixed point in Eq.~\eqref{eq:qedgn-fermionic} should be associated with the spontaneous breaking of TRS, and therewith represents the four-fermion version of the critical point in the original \qedgn\ theory, Eq.~\eqref{eq:lagrangian}.
At the c-QED$_3$ fixed point, we find at large $N$
\begin{align}
[\bar\psi \psi]_\text{c-QED$_3$} & =  2 + \mathcal O(1/N) =
[\bar\psi \sigma^z \psi]_\text{c-QED$_3$},
\end{align}
consistent with known results~\cite{chester2016}.
We remark that the $\mathcal O(1/N)$ corrections for the two operators are different~\cite{Hermele2007}.


\section{\texorpdfstring{\qedgn}{QED3-GN} quantum critical point in \texorpdfstring{$4-\epsilon$}{4-epsilon} expansion} \label{sec:eps-expansion}

The above one-loop calculation in the four-fermion theory space spanned by $u$ and $v$ allows us to obtain a qualitative picture of the structure of the RG flow, and to make contact with the situation in plain QED$_3$, when the order-parameter field $\phi$ is decoupled. However, in the physically interesting low-$N$ limit, the fixed points are located at strong coupling in $2+1$ dimensions, and the approximation ceases to be under perturbative control. One may therefore wonder whether it is possible to establish the existence and investigate the nature of the \qedgn\ fixed point within a complementary approach.
This is the subject of the present section.
To this end, we turn back to our initial formulation of the theory in terms of $\mathcal L_{\psi\phi}$, Eq.~\eqref{eq:lagrangian}. The Lagrangian can be generalized to arbitrary space-time dimension $2<D<4$ by trading the $2N$ flavors of two-component spinors for $N$ flavors of four-component spinors, and employing a $4\times 4$ representation of the Dirac matrices.
There are different possibilities on how to dimensionally continue the Dirac structure to noninteger $D$~\cite{dipietro2016}. Here, we use the common prescription that fixes the form of the TRS-breaking fermion bilinear $\bar\psi\psi$ in all $2<D<4$, as commonly done in the plain Gross-Neveu-Yukawa models~\cite{rosenstein1993,karkkainen1994,herbut2009,mihaila2017}.
In general $D$, the couplings have engineering dimensions
\begin{align}
[e^2] & = 4-D, &
[g] & = \frac{4-D}{2}, &
[\lambda] & = 4-D.
\end{align}
Hence, all three couplings \emph{simultaneously} become marginal when $D \nearrow 4$. This observation suggests that the \qedgn\ fixed point may be accessible perturbatively within an $\epsilon$ expansion near four space-time dimensions.

\begin{figure}[b]
\includegraphics[scale=.57]{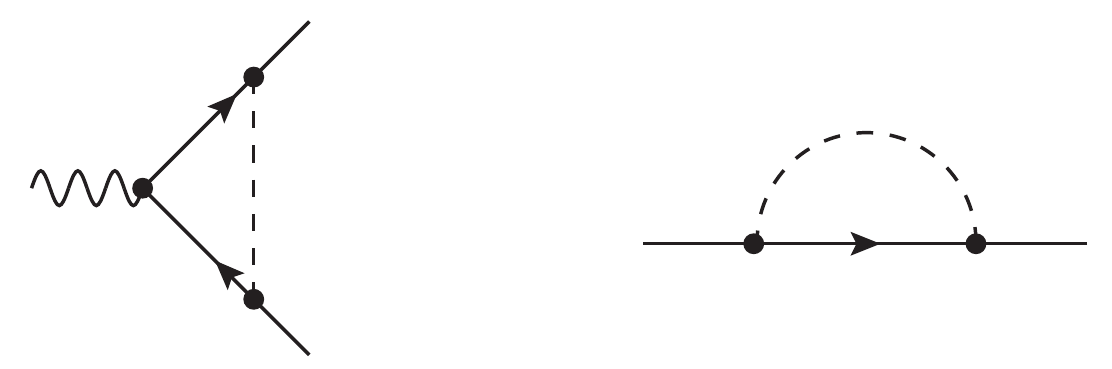}
\caption{Diagrams that cancel in the flow equation for the charge $e^2$ due to the Ward identity. In the flow equation for the Yukawa coupling $g^2$, the contribution from the fermion selfenergy (right) cancels with  the gauge-dependent part of the vertex correction (first diagram in Fig.~\ref{fig:scalar}).}
\label{fig:ward}
\end{figure}

In this limit, we find the flow equations for the couplings $e^2$, $g$, and $\lambda$ to the one-loop order as
\begin{align}
 \frac{d e^2}{d \ln b} & = (\epsilon - \eta_a) e^2, \\
 \frac{d g^2}{d \ln b} & = (\epsilon - \eta_\phi) g^2 + 6 e^2 g^2 - 3 g^4, \\
 \frac{d \lambda}{d \ln b} & = (\epsilon - 2\eta_\phi) \lambda - 36 \lambda^2 + N g^4,
\end{align}
with the anomalous dimensions
\begin{align}
 \eta_a & = \frac{4}{3} N e^2, & 
 \eta_\phi & = 2 N g^2,
\end{align}
where $N$ is the number of four-component fermions and $\epsilon = 4-D$. Here, we have tuned the system to criticality with $r \equiv 0$, and have rescaled $e^2/(8\pi^2) \mapsto e^2$, $g^2/(8\pi^2) \mapsto g^2$, and $\lambda/(8\pi^2) \mapsto \lambda$.
The corresponding diagrams are shown in Figs.~\ref{fig:gauge}, \ref{fig:scalar}, and \ref{fig:ward}.
Note that any dependence on the gauge-fixing parameter $\xi$ in the beta functions has canceled out, as it should be.
For $e^2 = 0$, the flow equations for $g^2$ and $\lambda$ coincide with those for the ungauged Gross-Neveu-Yukawa model~\cite{janssen2014}. For $g^2 = \lambda = 0$, on the other hand, the flow equation for the charge agrees with the one for QED$_{4-\epsilon}$~\cite{dipietro2016}. 

\begin{figure}[tb]
 \includegraphics[width=0.8\linewidth]{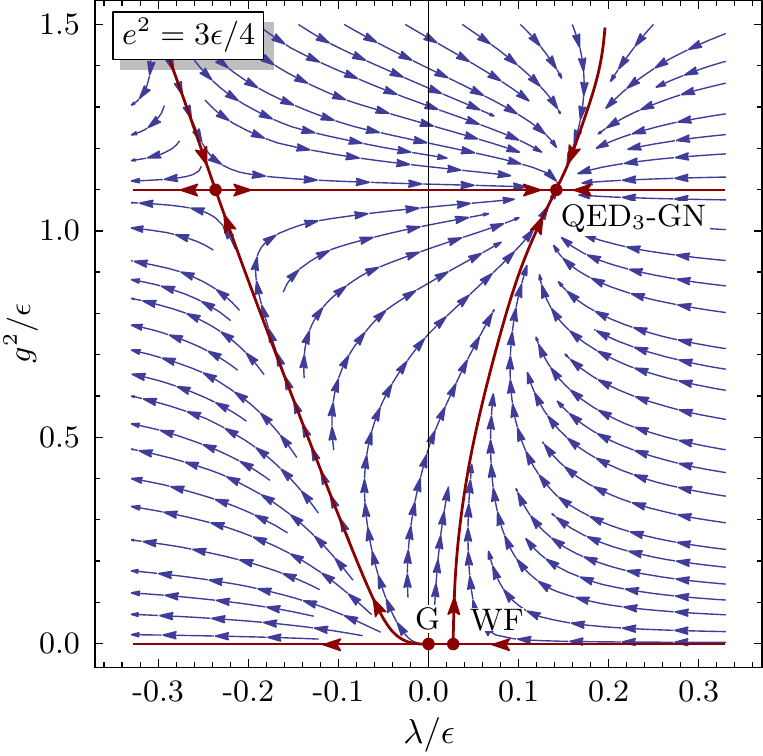}
 \caption{RG flow for $N=1$ in $(\lambda, g^2)$ plane to leading order in $\epsilon = 4-D$. For visualization purposes, here we have put the charge to its infrared fixed-point value $e^2 = e^2_* = 3\epsilon/4$, and we tune the system to criticality with $r\equiv 0$. The infrared stable fixed point at $g^2_* > 0$ and $\lambda_* > 0$ is the \qedgn\ quantum critical point and governs the transition into the TRS-broken state with $\langle \phi \rangle \neq 0$. G and WF at $g=0$ describe the Gaussian and Wilson-Fisher fixed points.}
 \label{fig:RG-flow-GNY}
\end{figure}

In the full theory space spanned by $e^2$, $g$, and $\lambda$, the above equations exhibit a unique infrared-stable fixed point at
\begin{align*}
 \text{\qedgn}&:& (e_*^2, g^2_*, \lambda_*) & = 
 \left(\tfrac{3}{4N}, \tfrac{2N+9}{2N(2N+3)}, \right. 
 \nonumber \\ &&& \quad 
 \left. \tfrac{-2N^2-15N+f(N)}{72N(2N+3)} \right)\epsilon + \mathcal O(\epsilon^2),
\end{align*}
where $f(N) \equiv \sqrt{4N^4+204N^3+1521N^2+2916N}$.
The fixed-point structure in the plane spanned by $\lambda$ and $g^2$ is illustrated for $N=1$ in Fig.~\ref{fig:RG-flow-GNY}. For visualization purposes, there we have set the charge $e^2$ to its infrared fixed-point value $e^2_*$. The \qedgn\ fixed point governs the continuous transition into the TRS-broken state with $\langle\phi\rangle \neq 0$, and should be understood as the Hubbard-Stratonovich-transformed version of the \qedgn\ fixed point we have found in the fermionic language, Eq.~\eqref{eq:qedgn-fermionic}. 
This is in full analogy to the equivalence of the critical points in the Gross-Neveu and Gross-Neveu-Yukawa theories~\cite{zinnjustin1991}.

\begin{figure*}[t]
 \includegraphics[width=0.325\textwidth]{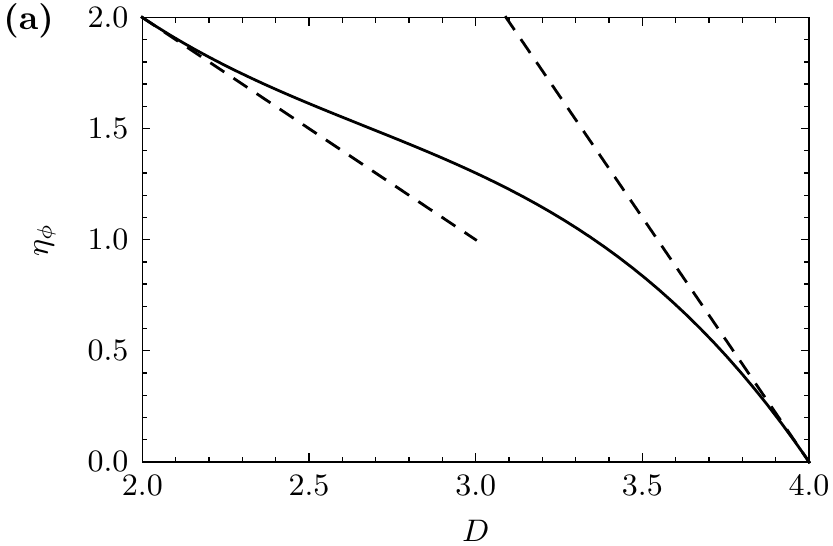}\hfill
 \includegraphics[width=0.325\textwidth]{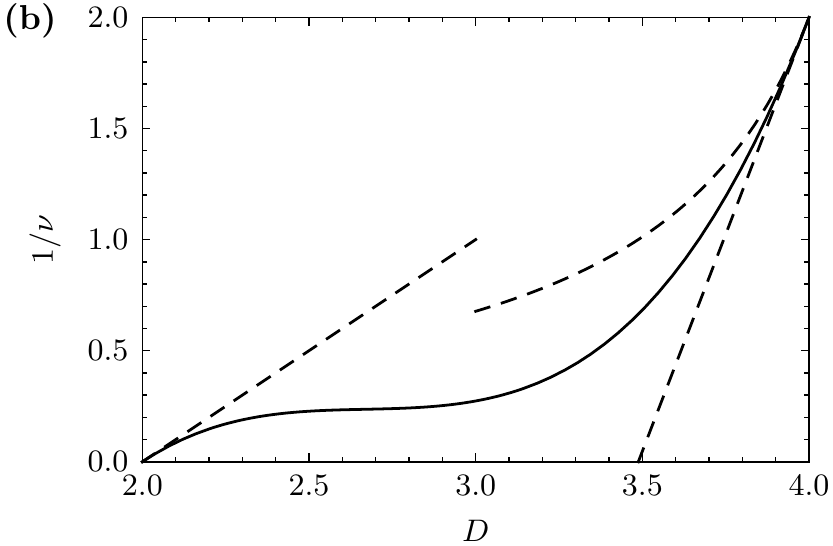}\hfill
  \includegraphics[width=0.325\textwidth]{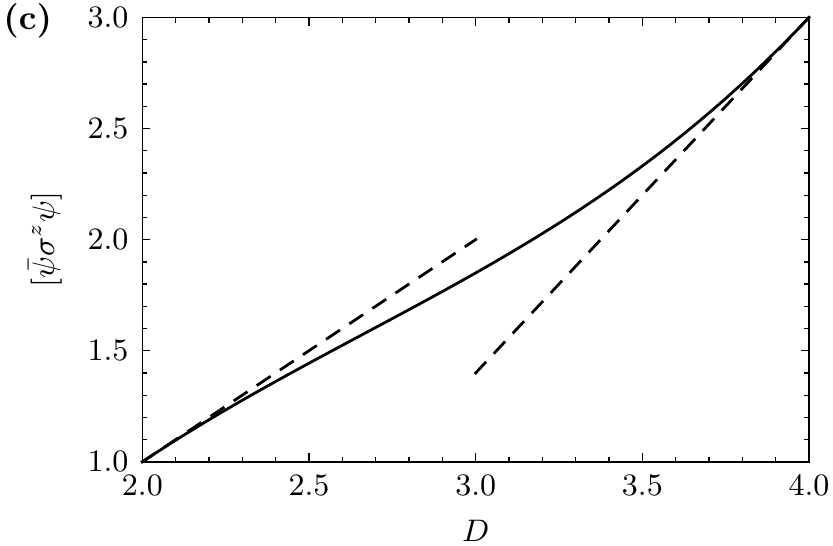}
 \caption{Critical exponents $\eta_\phi$ (a) and $1/\nu$ (b), and scaling dimension $[\bar\psi\sigma^z\psi]$ (c) as function of space-time dimension $D$ in the critical \qedgn\ theory for $N=1$. Solid curves: polynomial interpolation between lowest-order $2+\epsilon$ expansion result and $4-\epsilon$ expansion result. Dashed curves near lower and upper critical dimensions, respectively: plain extrapolation of $\epsilon$ expansion for comparison. For $1/\nu$ (b), the two dashed curves near $D=4$ correspond to the inverse of Eq.~\eqref{eq:nu-4-eps} (upper curve) and the expansion of $1/\nu$ itself (lower curve), cf.\ Ref.~\cite{janssen2014}.}
 \label{fig:eta-nu-interpolation}
\end{figure*}

In $D=2+1$ dimensions, the \qedgn\ fixed point is characterized by Lorentz invariance and $\mathrm{U}(2N)$ flavor symmetry. In order to be relevant for real materials, these symmetries must be \emph{emergent} in the low-energy limit. Flavor-symmetry-breaking perturbations have previously been shown to be indeed RG irrelevant, at least near the ungauged version of the fixed point (GN)~\cite{gehring2015}. Here, we demonstrate that the Lorentz symmetry also emerges in the critical region. As long as the spatial spherical symmetry remains intact, the only potentially relevant symmetry-breaking perturbations are terms quadratic in the fields. Adding these perturbations is equivalent to allowing different fermion and boson velocities, $v_\mathrm{F}$ and $v_\mathrm{B}$. Thus, we replace the kinetic terms in Eq.~\eqref{eq:lagrangian} by
\begin{align}
\gamma_\mu D_\mu & \mapsto \gamma_0 D_0 + v_\mathrm{F} \vec \gamma \cdot \vec D, &
\partial_\mu\partial_\mu & \mapsto \partial_0^2 + v_\mathrm{B}^2 {\vec\nabla}^2,
\end{align}
where $(D_\mu) \equiv (D_0, \vec D) \equiv  (\partial_\mu - i a_\mu)$, $\mu=0,\dots,D-1$, is the gauge-covariant derivative. $v_\mathrm{F}$ and $v_\mathrm{B}$ are measured in units of the speed of light $c\equiv 1$. Lorentz invariance is emergent when both flow to unity in the infrared, $v_{\mathrm{F,B}} \to 1$. The Lorentz-invariant subspace itself is invariant under the RG for symmetry reason. 
Allowing small symmetry-breaking perturbations out of this subspace as $v_\mathrm{F} = 1 + \delta v_\mathrm{F}$ and $v_\mathrm{B} = 1 + \delta v_\mathrm{B}$ with $\delta v_\mathrm{F,B} \ll 1$, we find the flow equations
\begin{align}
 \frac{d \delta v_\mathrm{F}}{d \ln b} & = - \frac{8e^2 + g^2}{3} \delta v_\mathrm{F} + \frac{g^2}{3} \delta v_\mathrm{B}, \\
 \frac{d \delta v_\mathrm{B}}{d \ln b} & = 2N g^2 \delta v_\mathrm{F} - 2 N g^2 \delta v_\mathrm{B}.
\end{align}
The corresponding stability matrix $\frac{\partial (d \delta v_\mathrm{F,B}/d \ln b)}{\partial \delta v_\mathrm{F,B}}$ has the eigenvalues
\begin{align}
 \theta_\pm = - \alpha \pm \sqrt{\alpha^2 - \beta^2}, 
\end{align}
with $\alpha \equiv (N+\frac{1}{6})g^2 + \frac{4}{3}e^2 > 0$ and $\beta^2 \equiv \frac{16}{3}Ne^2g^2 > 0$. Consequently, we have $\mathrm{Re}\,\theta_\pm < 0$ everywhere and $\delta v_\mathrm{F}$ and $\delta v_\mathrm{B}$ are always irrelevant perturbations. The Lorentz symmetry is therefore emergent at low energy.
This result is analogous to previous findings in related models with different order-parameter fields~\cite{roy2016}.

At the stable \qedgn\ fixed point, the anomalous dimensions read, to the leading order in $\epsilon = 4-D$,
\begin{align}
 \eta_{a} & = \epsilon, \label{eq:eta-a-4-eps}\\
 \eta_{\phi} & = \frac{2N + 9}{2N + 3}\epsilon + \mathcal O(\epsilon^2). \label{eq:eta-phi-4-eps}
\end{align}
We mention in passing that Eq.~\eqref{eq:eta-a-4-eps} is expected to not receive higher-order corrections due to the Ward identity associated with the U(1) gauge symmetry~\cite{herbut1996,janssen2016}.
The correlation-length exponent is related to the scaling dimension of $\phi^2$ via $1/\nu = D - [\phi^2]$. We obtain
\begin{align}
 \nu & = \frac{1}{2} + \frac{10N^2+39N+f(N)}{24N(2N+3)}\epsilon + \mathcal O(\epsilon^2). \label{eq:nu-4-eps}
 \end{align}
From the viewpoint of the duality conjecture, it is interesting to also compute the scaling dimension of the flavor-symmetry breaking bilinear $\bar\psi\sigma^z\psi$ at the \qedgn\ fixed point. To the leading order, we find
\begin{align}
 [\bar\psi\sigma^z\psi] & = 3 - \frac{2N+6}{2N+3}\epsilon + \mathcal O(\epsilon^2). \label{eq:dim-flavor-bilinear}
\end{align}

Now, if we simply extrapolated Eqs.~\eqref{eq:eta-a-4-eps}--\eqref{eq:dim-flavor-bilinear} towards large values of $\epsilon$, the leading-order corrections become sizable, e.g., $\eta_\phi \simeq 2.2\epsilon$, $\nu \simeq 0.5 + 0.98\epsilon$, and $[\bar\psi\sigma^z\psi] \simeq 3 - 1.6\epsilon$ for $N=1$. This obviously compromises the validity of the plain extrapolation.
The qualitative behavior of the exponents at large $\epsilon$ can, however, be inferred from the behavior near the \emph{lower} critical space-time dimension of two. From a calculation analogous to that leading to Eqs.~\eqref{eq:nu-fermionic}--\eqref{eq:eta-fermionic}, we find, to the lowest order,
\begin{align}
 1/\nu & = (D-2) + \mathcal O\!\left(1/N, (D-2)^2\right), \\
 \eta_\phi & = 2 - (D-2) + \mathcal O\!\left(1/N, (D-2)^2\right),
\end{align}
and
\begin{align}
 [\bar\psi\sigma^z\psi] & = 1 + (D-2) + \mathcal O\!\left(1/N, (D-2)^2\right).
\end{align}
This leading-order result coincides with the behavior of the plain Gross-Neveu model near the lower critical dimension~\cite{janssen2014, gracey2016}, which can be attributed to the fact that the charge contribution to the flow of $\Delta$ is subleading in $1/N$, c.f.\ Eq.~\eqref{eq:flow-Delta}.
In order to gain a reasonable estimate for the exponents in the physical situation in $D=3$ and small $N$, we can thus search for a smooth interpolation between the boundary values near the upper and lower critical dimensions. We use a simple polynomial form as in Ref.~\onlinecite{janssen2014}, and therewith find, for $D=3$,
\begin{align} \label{eq:eta-nu-d-3}
 \eta_\phi & \approx \frac{4N + 9}{2(2N + 3)}, &
 1/\nu & \approx \frac{50 N^2 + 51 N - f(N)}{24 N (2 N + 3)},
 \end{align}
 and
 \begin{align} \label{eq:psisigmapsi-d-3}
 [\bar\psi\sigma^z\psi] & \approx \frac{16N+21}{8N+12}.
\end{align}
The interpolating polynomials together with the naive extrapolations are depicted for $N=1$ as function of space-time dimension $2<D<4$ in Fig.~\ref{fig:eta-nu-interpolation}.
In the large-$N$ limit, Eqs.~\eqref{eq:eta-nu-d-3} and \eqref{eq:psisigmapsi-d-3} agree with the corresponding values calculated in Sec.~\ref{sec:fermionic}. For small $N$, the differences between Eqs.~\eqref{eq:eta-nu-d-3} and \eqref{eq:psisigmapsi-d-3} and the naivve extrapolations [Eqs.~\eqref{eq:eta-phi-4-eps}--\eqref{eq:dim-flavor-bilinear} for $\epsilon = 1$] can be viewed as a rough estimate on the accuracy of our results. For $N=1$, this gives $\eta_\phi \approx 1.3(9)$, $1/\nu \approx 0.3(4)$, and $[\bar\psi\sigma^z\psi] \approx 1.8(5)$.


\section{Conclusions} \label{sec:conclusions}

In this paper, we have studied the critical behavior of the \qedgn\ model in three space-time dimensions.
Just as in the corresponding plain Gross-Neveu universality class without a gauge field, there is a unique stable fixed point, which can be understood either as an ultraviolet fixed point of the four-fermion (``Gross-Neveu'') theory, or as an infrared fixed point of the partially bosonized (``Gross-Neveu-Yukawa'') theory~\cite{zinnjustin1991}.

We have employed the four-fermion language to clarify the correspondence of the \qedgn\ fixed point with the previously-studied conformal fixed point of plain QED$_3$~\cite{kubota2001,braun2014,dipietro2016,janssen2016,herbut2016}. 
Using this formulation, we have verified that the fixed-point annihilation mechanism that destabilizes the conformal phase of plain QED$_3$ at a critical flavor number $N_\mathrm{c}$, does not intrude upon the stability of the \qedgn\ fixed point. In fact, the latter turned out to continue to exist across the transition at $N_\mathrm{c}$ all the way down to $N=1$, at least within the present one-loop approximation.

The equivalent partially bosonized \qedgn\ theory, with a Yukawa interaction instead of the four-fermion term, can be dimensionally continued to noninteger space-time dimension $D$. We have used the fact that all three couplings present in the theory become simultaneously marginal when $D\nearrow 4$ to set up an $\epsilon$ expansion around four space-time dimensions. This allows to establish the existence of the \qedgn\ fixed point and to access the critical behavior in a controlled way.
We have computed the critical exponents $\eta_\phi$ and $\nu$, the scaling dimension of the flavor-symmetry-breaking bilinear $\bar\psi\sigma^z\psi$, as well as the gauge anomalous dimension $\eta_a$ to the leading order in $\epsilon=4-D$. 
For the latter, we predict $\eta_a = 4-D$ for all $N$ and $2<D<4$ exactly, which follows as a consequence of the Ward identity associated with the $\mathrm{U}(1)$ gauge symmetry. 
For the other exponents, our best estimates for $D=3$ are $\eta_\phi \approx 1.3(9)$ and $1/\nu \approx 0.3(4)$ in the case of $N=1$. Here, we have taken the difference between the plain extrapolation and the polynomial interpolation, which makes use of additional information of the behavior of the exponents near the lower critical dimension, as a rough error estimate.
The uncertainty becomes smaller for larger $N$, but for $N=1$ it is significantly larger than the error of the corresponding leading-order estimates in the plain Gross-Neveu universality class~\cite{janssen2014}. 
It would therefore be desirable to extend our work to higher loop order, e.g., along the lines carried out recently for the ungauged Gross-Neveu-Yukawa model~\cite{mihaila2017}.
As a complementary approach, the \qedgn\ fixed point should be accessible within the four-fermion formulation in an expansion around the \emph{lower} critical space-time dimension of two. The analogous computation in the plain Gross-Neveu model has now been accomplished, in a technological \emph{tour de force}, up to the four-loop order~\cite{gracey2016}. This necessitates to deal with the notorious evanescent operators, which render the theory nonunitary in dimensional regularization and are generically generated at high order in the $\epsilon$ expansion or when operators of high scaling dimension are analyzed~\cite{hogervorst2016}.

The comparatively large uncertainty of our results notwithstanding, we consider our finding of a large order-parameter anomalous dimension of order unity or larger to be reliable. In fact, a large value of $\eta_\phi$ appears to be characteristic to all known chiral universality classes that are driven by massless fermionic degrees of freedom~\cite{assaad2013,chandrasekharan2013,toldin2015,li2015,otsuka2016}. Theoretically, this property can be traced back to the observation that in all critical fermion systems the order-parameter anomalous dimension has to approach unity in the limit of large flavor number. Furthermore, near the lower critical dimension, its boundary value is $\eta_\phi = 2 + \mathcal O(D-2)$. 

\begin{table}[t]
\caption{Scaling dimensions of operators at the $N=1$ \qedgn\ fixed point in comparison with literature values for scaling dimensions of the corresponding dual operators at the N\'eel-VBS deconfined critical point~\cite{sandvik2007,melko2008,nahum2015a}. The latter is presumably described by the SU(2) \nccp\ model, for which we also quote the results of a field-theoretical approach~\cite{bartosch2013}.}
\label{tab:exponents}
\begin{tabular*}{\linewidth}{@{}rl@{\extracolsep{\fill} }r@{\extracolsep{0pt} }lr@{}}
\hline\hline
\multicolumn{2}{c}{\qedgn} & \multicolumn{3}{c}{SU(2) \nccp} \\
\hline
$[\phi] \approx $ & $(1 + 1.3(9))/2$ & $[z^\dagger \sigma^z z] \approx $ & $(1 + 0.26(3))/2$ & \cite{sandvik2007} \\
&& $\approx$ & $(1+0.35(3))/2$ & \cite{melko2008} \\
&& $\approx$ & $(1+0.25(3))/2$ & \cite{nahum2015a} \\
&& $\approx$ & $(1+0.22)/2$ & \cite{bartosch2013}\\
$[\bar\psi\sigma^z\psi] \approx $ & $3 - 1.2(5)$ & $[z^\dagger z] \approx $ & $3 - 1.28(5)$& \cite{sandvik2007}\\
&& $\approx$ & $3 - 1.47(9)$ & \cite{melko2008}\\
&& $\approx$ & $3 - 1.99(4)$ & \cite{nahum2015a}\\
&& $\approx$ & $3 - 1.79$ & \cite{bartosch2013}\\
$[\phi^2] \approx $ & $3 - 0.3(4)$ & $[z^\dagger z]\approx $ & \multicolumn{2}{c}{\emph{---see above---}} \\
\hline\hline
\end{tabular*}
\end{table}

These findings are striking in the light of the recently conjectured duality of the $N=1$ \qedgn\ theory with the $\mathrm{SU}(2)$ \nccp\ model~\cite{wang2017}, which in turn is believed to describe the deconfined critical point between the N\'eel and VBS phases of spin-$1/2$ systems on the square lattice~\cite{senthil2004a,senthil2004b,nahum2015a,nahum2015b}. While the existence of a stable \qedgn\ fixed point is a prerequisite for the duality scenario to hold, our leading-order results for its critical behavior is not entirely compatible with the critical (or pseudocritical) behavior measured in the spin systems. The largest discrepancy occurs in the case of the order-parameter anomalous dimensions, which in the spin systems have been determined as $\eta_\text{N\'eel} \approx \eta_\text{VBS} \approx 0.25 \dots 0.35$~\cite{sandvik2007,melko2008,nahum2015a}. This is about an order of magnitude larger than in the standard bosonic $\mathrm{O}(5)$ universality class~\cite{hasenbusch2005}, but still significantly smaller than our estimate of $\eta_\phi \approx 1.3(9)$ in the \qedgn\ theory. Direct simulations of the \nccp\ model remain inconclusive as to whether the transition is continuous~\cite{motrunich2004,motrunich2008,wellegehausen2014} or weakly first order~\cite{kuklov2008}. 
In any case, as far as we are aware, at present no numerical data in the purely bosonic models appear to suggest an anomalous dimension of order unity or larger.
Field-theoretical approaches to the critical behavior of the \nccp\ model appear to be difficult, since the loop corrections are sizable~\cite{irkhin1996,kaul2008}. Nevertheless, a functional RG approach finds values that are remarkably close to the most recent numerical results in the spin systems~\cite{bartosch2013}.

We have also computed the scaling dimension of the flavor-symmetry-breaking fermion bilinear $\bar\psi\sigma^z\psi$, which is identified with $z^\dagger z$ in the bosonic \nccp\ theory. The latter corresponds to the tuning parameter for the N\'eel-VBS transition. Therefore, the duality predicts  $1/\nu_\text{N\'eel-VBS}=3- [\bar\psi\sigma^z\psi]$.
Our calculation gives $3-[\bar\psi\sigma^z\psi] \approx 1.2(5)$, while the numerical simulations of the N\'eel-VBS transition find $1/\nu_\text{N\'eel-VBS} \approx 1.3 \dots 2.0$~\cite{sandvik2007,melko2008,nahum2015a}. These values are not inconsistent with the duality prediction.
The duality also predicts that the scaling dimension $[\bar\psi\sigma^z \psi]$ should coincide with $[\phi^2]$.
Our result for $\nu$ gives $[\phi^2]=3-1/\nu\approx 2.7(4)$ which is only somewhat larger than $[\bar\psi\sigma^z \psi] \approx 1.8(5)$, but incompatible with the numerical ranges quoted for $3 - 1/\nu_\text{N\'eel-VBS}$.

In conclusion, our estimate for $[\bar\psi \sigma^z \psi]$ seems to be not inconsistent with the duality proposal, but $[\phi]$ and $[\phi^2]$ show large discrepancies when comparing them with the corresponding measurements in the bosonic systems. This is summarized in Table~\ref{tab:exponents}.
We note, however, that if the transition in the spin systems is indeed continuous with an emergent $\mathrm{SO}(5)$ symmetry~\cite{nahum2015b}, then this unavoidably necessitates anomalous dimensions that are significantly above the ones currently observed~\cite{nakayama2016}.
We therefore believe that the possibility that higher-order computations in the \qedgn\ model and forthcoming numerical calculations in the spin systems converge to common values in future works is as yet not excluded.
%


\acknowledgments
We thank S. Bhattacharjee, S. Rychkov, and C. Wang for helpful discussions. L.\,J.\ acknowledges support by the DFG under SFB1143 ``Correlated Magnetism: From Frustration to Topology.'' Y.-C.\,H.\ is supported by the Gordon and Betty Moore Foundation under the EPiQS initiative, GBMF4306, at Harvard University.




\begin{thebibliography}{100}

\bibitem{sachdevbook}
S. Sachdev,
Quantum phase transition,
Cambridge University Press, (2011)

\bibitem{senthil2004a}
T. Senthil, A. Vishwanath, L. Balents, S. Sachdev, and M. P. A. Fisher,
Deconfined quantum critical points,
Science {\bf 303}, 1490 (2004).

\bibitem{senthil2004b} 
T. Senthil, L. Balents, S. Sachdev, A. Vishwanath, and M. P. A. Fisher,
Quantum criticality beyond the Landau-Ginzburg-Wilson paradigm,
Phys. Rev. B {\bf 70}, 144407 (2004).

\bibitem{Grover2013}
T. Grover and A. Vishwanath,
Quantum phase transition between integer quantum Hall states of bosons,
Phys. Rev. B {\bf 87}, 045129,  (2013).

\bibitem{Lu2014}
Y.-M. Lu and D.-H. Lee,
Quantum phase transitions between bosonic symmetry-protected topological phases in two dimensions: Emergent 
QED$_3$  and anyon superfluid,
Phys. Rev. B {\bf 89}, 195143, (2014).

\bibitem{he2015b}
Y.-C. He, Y. Fuji, and S. Bhattacharjee, 
Kagome spin liquid: a deconfined critical phase driven by U(1) gauge fluctuation,
arXiv:1512.05381.

\bibitem{Kalmeyer1987}
V. Kalmeyer and R. B. Laughlin,
Equivalence of the resonating-valence-bond and fractional quantum Hall states,
Phys. Rev. Lett. {\bf 59}, 2095 (1987).

\bibitem{Wen1989}
X. G. Wen, F. Wilczek, and A. Zee,
Chiral spin states and superconductivity,
Phys. Rev. B {\bf 39}, 11413 (1989).

\bibitem{Hastings2000}
M. B. Hastings,
Dirac structure, RVB, and Goldstone modes in the kagome antiferromagnet,
Phys. Rev. B {\bf 63}, 014413  (2000).

\bibitem{Chen2013}
X. Chen, Z.-C. Gu, Z.-X. Liu, and X.-G. Wen,
Symmetry protected topological orders and the group cohomology of their symmetry group,
Phys. Rev. B {\bf 87}, 155114 (2013).

\bibitem{sato2017}
T. Sato, M. Hohenadler, and F. F. Assaad,
Dirac Fermions with Competing Mass Terms: Non-Landau Transition with Emergent Symmetry,
arXiv:1707.03027.

\bibitem{motrunich2004}
O. I. Motrunich and A. Vishwanath,
Emergent photons and transitions in the O(3) sigma model with hedgehog suppression,
Phys. Rev. B {\bf 70}, 075104 (2004).

\bibitem{peskin1978}
M. E. Peskin,
Mandelstam-'t Hooft duality in Abelian Lattice Models,
Ann. Phys. {\bf 113}, 122 (1978).

\bibitem{dasgupta1981}
C. Dasgupta and B. I. Halperin, 
Phase Transition in a Lattice Model of Superconductivity,
Phys. Rev. Lett. {\bf 47}, 1556 (1981).

\bibitem{fisher1989}
P. A. Fisher and D. H. Lee, 
Correspondence between two-dimensional bosons and a bulk superconductor in a magnetic field,
Phys. Rev. B {\bf 39}, 2756 (1989).

\bibitem{son2015}
D. T. Son, 
Is the Composite Fermion a Dirac Particle?,
Phys. Rev. X {\bf 5}, 031027 (2015).

\bibitem{wang2015}
C. Wang and T. Senthil, 
Dual Dirac Liquid on the Surface of the Electron Topological Insulator,
Phys. Rev. X {\bf 5}, 041031 (2015).

\bibitem{metlitski2016}
M. A. Metlitski and A. Vishwanath, 
Particle-Vortex Duality of Two-Dimensional Dirac Fermion from Electric-Magnetic Duality of Three-Dimensional Topological Insulators,
Phys. Rev. B {\bf 93}, 245151 (2016).

\bibitem{mross2016}
D. F. Mross, J. Alicea, and O. I. Motrunich, 
Explicit Derivation of Duality Between a Free Dirac Cone and Quantum Electrodynamics in (2+1) Dimensions,
Phys. Rev. Lett. {\bf 117}, 016802 (2016).

\bibitem{xu2015}
C. Xu and Y.-Z. You, 
Self-dual quantum electrodynamics as boundary state of the three-dimensional bosonic topological insulator,
Phys. Rev. B {\bf 92}, 220416(R) (2015).

\bibitem{seiberg2016}
N. Seiberg, T. Senthil, C. Wang, and E. Witten,
A duality web in 2+1 dimensions and condensed matter physics,
Ann. Phys. {\bf 374}, 395 (2016).

\bibitem{karch2016}
A. Karch and D. Tong,
Particle-Vortex Duality from 3D Bosonization,
Phys. Rev. X {\bf 6}, 031043 (2016).

\bibitem{hsien2016}
P.-S. Hsin and N. Seiberg,
Level/rank duality and Chern-Simons-matter theories,
J. High Energy Phys. 09 (2016), 95.

\bibitem{murugan2017}
J. Murugan and N. Horatiu,
Particle-vortex duality in topological insulators and superconductors,
J. High Energy Phys. 05 (2017), 1.

\bibitem{Mross2017}
D. F. Mross, J. Alicea, and O. I. Motrunich,
Symmetry and duality in bosonization of two-dimensional Dirac fermions,
arXiv:1705.01106 (2017).
 
\bibitem{Chen2017}
J.-Y. Chen, J. H. Son, C, Wang, and S. Raghu,
Exact Boson-Fermion Duality on a 3D Euclidean Lattice,
arXiv:1705.05841 (2017).

\bibitem{alicea2005}
J. Alicea, O. I. Motrunich, M. Hermele, and M. P. A. Fisher,
Criticality in quantum triangular antiferromagnets via fermionized vortices,
Phys. Rev. B {\bf 72}, 064407 (2005).

\bibitem{senthil2006}
T. Senthil and M. P. A. Fisher, 
Competing orders, nonlinear sigma models, and topological terms in quantum magnets,
Phys. Rev. B {\bf 74}, 064405 (2006).

\bibitem{wang2017}
C. Wang, A. Nahum, M. A. Metlitski, C. Xu, and T. Senthil,
Deconfined quantum critical points: symmetries and dualities,
arXiv:1703.02426.

\bibitem{Qin2017}
Y. Q. Qin, Y.-Y He, Y.-Z. You, Z.-Y Lu, A. Sen, A. W. Sandvik, C. Xu, and Z. Y. Meng,
Duality between the deconfined quantum-critical point and the bosonic topological transition,
arXiv:1705.10670.

\bibitem{nahum2015a}
A. Nahum, J. T. Chalker, P. Serna, M. Ortu\~{n}o, and A. M. Somoza,
Deconfined Quantum Criticality, Scaling Violations, and Classical Loop Models,
Phys. Rev. X {\bf 5}, 041048 (2015).

\bibitem{nahum2015b}
A. Nahum, P. Serna, J. T. Chalker, M. Ortu\~{n}o, and A. M. Somoza, 
Emergent SO(5) Symmetry at the N\'eel to Valence-Bond-Solid Transition,
Phys. Rev. Lett. {\bf 115}, 267203 (2015).

\bibitem{appelquist1988}
T. Appelquist, D. Nash, and L. C. R. Wijewardhana,
Critical Behavior in (2+1)-Dimensional QED,
Phys. Rev. Lett. {\bf 60}, 2575 (1988).

\bibitem{kubota2001}
K. Kubota and H. Terao,
Dynamical Symmetry Breaking in QED$_3$ from the Wilson RG Point of View,
Prog. Theor. Phys. {\bf 105}, 809 (2001).

\bibitem{braun2014}
J. Braun, H. Gies, L. Janssen, and D. Roscher,
Phase structure of many-flavor QED$_3$,
Phys. Rev. D {\bf 90}, 036002 (2014).

\bibitem{janssen2016}
L. Janssen,
Spontaneous breaking of Lorentz symmetry in $(2+\epsilon)$-dimensional QED,
Phys. Rev. D {\bf 94}, 094013 (2016).

\bibitem{herbut2016}
I. F. Herbut,
Chiral symmetry breaking in three-dimensional quantum electrodynamics as fixed point annihilation,
Phys. Rev. D {\bf 94}, 025036 (2016).

\bibitem{dipietro2016}
L. Di Pietro, Z. Komargodski, I. Shamir, and E. Stamou,
Quantum Electrodynamics in $d=3$ from the $\epsilon$ Expansion,
Phys. Rev. Lett. {\bf 116}, 131601 (2016).

\bibitem{chester2016}
S. M. Chester and S. S. Pufu,
Anomalous dimensions of scalar operators in QED$_3$,
J. High Energy Phys. 08 (2016), 1.

\bibitem{gusynin2016}
V. P. Gusynin and P. K. Pyatkovskiy,
Critical number of fermions in three-dimensional QED,
Phys. Rev. D {\bf 94}, 125009 (2016).
 
\bibitem{kotikov2016b}
A. V. Kotikov and S. Teber,
Critical behavior of $(2+1)$-dimensional QED: $1/N_f$ corrections in an arbitrary nonlocal gauge,
Phys. Rev. D {\bf 94}, 114011 (2016).

\bibitem{hands2004}
S. J. Hands, J. B. Kogut, L. Scorzato, and C. G. Strouthos,
Noncompact three-dimensional quantum electrodynamics with $N_f=1$ and $N_f=4$,
Phys. Rev. B {\bf 70}, 104501 (2004).

\bibitem{raviv2014}
O. Raviv, Y. Shamir, and B. Svetitsky,
Nonperturbative beta function in three-dimensional electrodynamics,
Phys. Rev. D {\bf 90}, 014512 (2014).

\bibitem{karthik2016a}
N. Karthik and R. Narayanan,
No evidence for bilinear condensate in parity-invariant three-dimensional QED with massless fermions,
Phys. Rev. D {\bf 93}, 045020 (2016).

\bibitem{karthik2016b}
N. Karthik and R. Narayanan,
Scale invariance of parity-invariant three-dimensional QED,
Phys. Rev. D {\bf 94}, 065026 (2016).

\bibitem{sandvik2007}
A. W. Sandvik,
Evidence for Deconfined Quantum Criticality in a Two-Dimensional Heisenberg Model with Four-Spin Interactions,
Phys. Rev. Lett. {\bf 98}, 227202 (2007).

\bibitem{melko2008}
R. G. Melko and R. K. Kaul,
Scaling in the Fan of an Unconventional Quantum Critical Point,
Phys. Rev. Lett. {\bf 100}, 017203 (2008).

\bibitem{gracey2016}
J. A. Gracey, T. Luthe, and Y. Schr\"oder,
Four loop renormalization of the Gross-Neveu model,
Phys. Rev. D {\bf 94}, 125028 (2016).

\bibitem{mihaila2017}
L. N. Mihaila, N. Zerf, B. Ihrig, I. F. Herbut, and M. M. Scherer, 
Gross-Neveu-Yukawa model at three loops and Ising critical behavior of Dirac systems,
arXiv:1703.08801.

\bibitem{janssen2014}
L. Janssen and I. F. Herbut,
Antiferromagnetic critical point on graphene's honeycomb lattice: A functional renormalization group approach,
Phys. Rev. B {\bf 89}, 205403 (2014).

\bibitem{heilmann2015}
M. Heilmann, T. Hellwig, B. Knorr, M. Ansorg, and A. Wipf, 
Convergence of Derivative Expansion in Supersymmetric Functional RG Flows,
J. High Energy Phys. 02 (2015), 109.

\bibitem{knorr2016}
B. Knorr, 
Ising and Gross-Neveu model in next-to-leading order,
Phys. Rev. B {\bf 94}, 245102 (2016).

\bibitem{iliesu2016}
L. Iliesiu, F. Kos, D. Poland, S. S. Pufu, D. Simmons-Duffin, and R. Yacoby, 
Bootstrapping 3D Fermions, 
J. High Energy Phys. 03 (2016), 120.

\bibitem{iliesu2017}
L. Iliesiu, F. Kos, D. Poland, S. S. Pufu, and D. Simmons-Duffin,
Bootstrapping 3D Fermions with Global Symmetries,
arXiv:1705.03484.

\bibitem{assaad2013}
F. F. Assaad and I. F. Herbut,
Pinning the Order: The Nature of Quantum Criticality in the Hubbard Model on Honeycomb Lattice,
Phys. Rev. X {\bf 3}, 031010 (2013).

\bibitem{chandrasekharan2013}
S. Chandrasekharan and A. Li, 
Quantum critical behavior in three dimensional lattice Gross-Neveu models,
Phys. Rev. D {\bf 88}, 021701 (2013).

\bibitem{toldin2015}
F. Parisen Toldin, M. Hohenadler, F. F. Assaad, and I. F. Herbut,
Fermionic quantum criticality in honeycomb and $\pi$-flux Hubbard models: Finite-size scaling of renormalization-group-invariant observables from quantum Monte Carlo,
Phys. Rev. B {\bf 91}, 165108 (2015).

\bibitem{li2015}
Z.-X. Li, Y.-F. Jiang, and H. Yao, 
Fermion-sign-free Majarana-quantum-Monte-Carlo studies of quantum critical phenomena of Dirac fermions in two dimensions,
New J. Phys. {\bf 17}, 085003 (2015).

\bibitem{otsuka2016}
Y. Otsuka, S. Yunoki, and S. Sorella,
Universal Quantum Criticality in the Metal-Insulator Transition of Two-Dimensional Interacting Dirac Electrons,
Phys. Rev. X {\bf 6}, 011029 (2016).

\bibitem{he2014}
Y.-C. He, D. N. Sheng, and Y. Chen,
Chiral Spin Liquid in a Frustrated Anisotropic Kagome Heisenberg Model,
Phys. Rev. Lett. {\bf 112}, 137202 (2014).

\bibitem{Gong2014}
S.‰-S. Gong, W. Zhu, and D. ‰N. Sheng, 
Emergent Chiral Spin Liquid: Fractional Quantum Hall Effect in a Kagome Heisenberg Model,
Sci. Rep. {\bf 4}, 6317 (2014).

\bibitem{he2015a}
Y.-C. He and Y. Chen,
Distinct Spin Liquids and their Transitions in Spin-1/2 XXZ Kagome Antiferromagnets
Phys. Rev. Lett. {\bf 114}, 037201 (2015).

\bibitem{he2016}
Y.-C. He, M. P. Zaletel, M. Oshikawa, and F. Pollmann,
Signatures of Dirac cones in a DMRG study of the Kagome Heisenberg model,
Phys. Rev. X {\bf7}, 031020 (2017).

\bibitem{Chen1993}
W. Chen, M. P. A. Fisher, and Y.-S. Wu, 
Mott transition in an anyon gas,
Phys. Rev. B {\bf 48}, 13749 (1993).

\bibitem{jackiw1981}
R. Jackiw and S. Templeton,
How super-renormalizable interactions cure their infrared divergences,
Phys. Rev. D {\bf 23}, 2291 (1981).

\bibitem{halperin1974}
B. I. Halperin, T. C. Lubensky, and S.-k. Ma,
First-Order Phase Transitions in Superconductors and Smectic-$A$ Liquid Crystals,
Phys. Rev. Lett. {\bf 32}, 292 (1974).

\bibitem{jaeckel2006}
H. Gies and J. Jaeckel,
Chiral phase structure of QCD with many flavors,
Eur. Phys. J. C {\bf 46}, 433 (2006).

\bibitem{kaplan2009}
D. B. Kaplan, J.-W. Lee, D. T. Son, and M. A. Stephanov,
Conformality lost,
Phys. Rev. D {\bf 80}, 125005 (2009).

\bibitem{herbut2014}
I. F. Herbut and L. Janssen,
Topological Mott Insulator in Three-Dimensional Systems with Quadratic Band Touching,
Phys. Rev. Lett. {\bf 113}, 106401 (2014).

\bibitem{janssen2017}
L. Janssen and I. F. Herbut,
Phase diagram of electronic systems with quadratic Fermi nodes in $2<d<4$: $2+\epsilon$ expansion, $4-\epsilon$ expansion, and functional renormalization group,
Phys. Rev. B {\bf 95}, 075101 (2017).

\bibitem{kragset2006}
S. Kragset, E. Sm\o{}rgrav, J. Hove, F. S. Nogueira, and A. Sudb\o{},
First-Order Phase Transition in Easy-Plane Quantum Antiferromagnets,
Phys. Rev. Lett. {\bf 97}, 247201 (2006).

\bibitem{emidio2016}
J. D'Emidio and R. K. Kaul,
First-order superfluid to valence-bond solid phase transitions in easy-plane SU($N$) magnets for small $N$,
Phys. Rev. B {\bf 93}, 054406 (2016).

\bibitem{emidio2017}
J. D'Emidio and R. K. Kaul,
New Easy-Plane $\mathbbm{CP}^{N-1}$ Fixed Points,
Phys. Rev. Lett. {\bf 118}, 187202 (2017).

\bibitem{Zhang2017}
X.-F. Zhang, Y.-C. He, S. Eggert, R. Moessner, and F. Pollmann,
Continuous easy-plane deconfined phase transition on the kagome lattice,
arXiv:1706.05414.

\bibitem{roy2013}
See, however,
B. Roy, V. Juri\v{c}i\'c, and I. F. Herbut,
Quantum superconducting criticality in graphene and topological insulators,
Phys. Rev. B {\bf 87}, 041401(R) (2013),
for the critical behavior of a related Gross-Neveu-Yukawa theory with a different order-parameter field.

\bibitem{yan2011}
S. Yan, D. A. Huse, and S. R. White, 
Spin-Liquid Ground State of the $S = 1/2$ Kagome Heisenberg Antiferromagnet,
Science {\bf 332}, 1173 (2011).

\bibitem{depenbrock2012}
S. Depenbrock, I. P. McCulloch, and U. Schollw\"ock, 
Nature of the Spin-Liquid Ground State of the $S = 1/2$ Heisenberg Model on the Kagome Lattice,
Phys. Rev. Lett. {\bf 109}, 067201 (2012).

\bibitem{jiang2012}
H.-C. Jiang, Z. Wang, and L. Balents, 
Identifying Topological Order by Entanglement Entropy,
Nature Phys. {\bf 8}, 902 (2012).

\bibitem{ran2007}
Y. Ran, M. Hermele, P. A. Lee, and X.-G. Wen,
Projected-Wave-Function Study of the Spin-$1/2$ Heisenberg Model on the Kagom\'e Lattice,
Phys. Rev. Lett. {\bf 98}, 117205 (2007).

\bibitem{iqbal2015}
Y. Iqbal, D. Poilblanc, and F. Becca,
Spin-$\frac{1}{2}$ Heisenberg $J_1$-$J_2$ antiferromagnet on the kagome lattice,
Phys. Rev. B {\bf 91}, 020402(R) (2015).

\bibitem{gies2010}
H. Gies and L. Janssen,
UV fixed-point structure of the three-dimensional Thirring model,
Phys. Rev. D {\bf 82}, 085018 (2010).

\bibitem{herbut2006}
I. F. Herbut,
Interactions and Phase Transitions on Graphene's Honeycomb Lattice,
Phys. Rev. Lett. {\bf 97}, 146401 (2006).

\bibitem{herbut2009}
I. F. Herbut, V. Juri\v{c}i\'{c}, and  O. Vafek,
Relativistic Mott criticality in graphene,
Phys. Rev. B {\bf 80}, 075432 (2009).

\bibitem{braun2011}
J. Braun, H. Gies, and D. D. Scherer,
Asymptotic safety: a simple example,
Phys. Rev. D {\bf 83} 085012 (2011).
 
\bibitem{janssen2012}
L. Janssen and H. Gies,
Critical behavior of the (2+1)-dimensional Thirring model,
Phys. Rev. D {\bf 86}, 105007 (2012).

\bibitem{christofi2007}
S. Christofi, S. Hands, and C. Strouthos,
Critical flavor number in the three dimensional Thirring model,
Phys. Rev. D {\bf 75}, 101701(R) (2007).

\bibitem{hands2016}
S. Hands,
Towards critical physics in 2+1$d$ with U(2$N$)-invariant fermions,
J. High Energy Phys. 11 (2016), 15.

\bibitem{wellegehausen2017}
B. H. Wellegehausen, D. Schmidt, and A. Wipf,
Critical flavour number of the Thirring model in three dimensions,
arXiv:1708.01160.

\bibitem{kaveh2005}
K. Kaveh and I. F. Herbut,
Chiral symmetry breaking in three-dimensional quantum electrodynamics in the presence of irrelevant interactions: A renormalization group study,
Phys. Rev. B {\bf 71}, 184519 (2005).

\bibitem{Hermele2007}
M. Hermele, T. Senthil, and M. P. A. Fisher, 
Algebraic Spin Liquid as the Mother of Many Competing Orders, 
Phys. Rev. B {\bf72}, 104404 (2005); Phys. Rev. B {\bf76}, 149906(E) (2007).

\bibitem{rosenstein1993}
B. Rosenstein, H.-L. Yu, and A. Kovner,
Critical exponents of new universality classes,
Phys. Lett. B {\bf 314}, 381 (1993).

\bibitem{karkkainen1994}
L. K\"{a}rkk\"{a}inen, R. Lacaze, P. Lacock, and B. Petersson,
Critical behaviour of the three-dimensional Gross-Neveu and Higgs-Yukawa models,
Nucl. Phys. B {\bf 415}, 781 (1994).

\bibitem{zinnjustin1991}
J. Zinn-Justin,
Four-fermion interaction near four dimensions,
Nucl. Phys. B {\bf 367}, 105 (1991).

\bibitem{gehring2015}
F. Gehring, H. Gies, and L. Janssen,
Fixed-point structure of low-dimensional relativistic fermion field theories: Universality classes and emergent symmetry,
Phys. Rev. D {\bf 92}, 085046 (2015).

\bibitem{roy2016}
B. Roy, V. Juri\v{c}i\'{c}, and I. F. Herbut,
Emergent Lorentz symmetry near fermionic quantum critical points in two and three dimensions,
J. High Energy Phys. 04 (2016), 18.

\bibitem{herbut1996}
I. F. Herbut and Z. Te\v{s}anovi\'{c},
Critical Fluctuations in Superconductors and the Magnetic Field Penetration Depth,
Phys. Rev. Lett. {\bf 76}, 4588 (1996).

\bibitem{hogervorst2016}
M. Hogervorst, S. Rychkov, and B. C. van Rees,
Unitarity violation at the Wilson-Fisher fixed point in $4-\epsilon$ dimensions,
Phys. Rev. D {\bf 93}, 125025 (2016).

\bibitem{hasenbusch2005}
M. Hasenbusch, A. Pelissetto, and E. Vicari,
Instability of O(5) multicritical behavior in SO(5) theory of high-$T_c$ superconductors,
Phys. Rev. B {\bf 72}, 014532 (2005).

\bibitem{motrunich2008}
O. I. Motrunich and A. Vishwanath, 
Comparative study of Higgs transition in one-component and two-component lattice superconductor models,
arXiv:0805.1494.

\bibitem{wellegehausen2014}
B. H. Wellegehausen, D. K\"orner, and A. Wipf,
Asymptotic safety on the lattice: The nonlinear sigma model,
Ann. Phys. {\bf 349}, 374 (2014).

\bibitem{kuklov2008}
A. B. Kuklov, M. Matsumoto, N. V. Prokof'ev, B. V. Svistunov, and M. Troyer,
Deconfined Criticality: Generic First-Order Transition in the SU(2) Symmetry Case,
Phys. Rev. Lett. {\bf 101}, 050405 (2008).

\bibitem{irkhin1996}
V. Y. Irkhin, A. A. Katanin, and M. I. Katsnelson,
$1/N$ expansion for critical exponents of magnetic phase transitions in the $CP^{N-1}$ model for $2<d<4$,
Phys. Rev. B {\bf 54}, 11953 (1996).

\bibitem{kaul2008}
R. K. Kaul and S. Sachdev,
Quantum criticality of U(1) gauge theories with fermionic and bosonic matter in two spatial dimensions,
Phys. Rev. B {\bf 77}, 155105 (2008).

\bibitem{bartosch2013}
L. Bartosch,
Corrections to scaling in the critical theory of deconfined criticality,
Phys. Rev. B {\bf 88}, 195140 (2013).

\bibitem{nakayama2016}
Y. Nakayama and T. Ohtsuki,
Necessary Condition for Emergent Symmetry from the Conformal Bootstrap,
Phys. Rev. Lett. {\bf 117}, 131601 (2016).

\end{thebibliography}
\end{document}